\documentclass{andp2012}
\usepackage[english]{babel}
\emergencystretch=\maxdimen
\title{Detecting non-Markovianity via linear entropy of Choi state}
\author[Xiao Zheng]{Xiao Zheng\inst{1}}
\author[Shao-Qiang Ma]{Shao-Qiang Ma\inst{1}}
\author[Guo-Feng Zhang]{Guo-Feng Zhang\inst{1,}\footnote{Corresponding author\quad E-mail:~\textsf{gf1978zhang@buaa.edu.cn}}}
\address[1]{Key Laboratory of Micro-Nano Measurement-Manipulation and Physics (Ministry of
Education), School of Physics, Beihang University, Xueyuan Road No. 37, Beijing 100191, China}
\shortauthors{X. Zheng et al.}
\begin{abstract}

Non-Markovian dynamics detection is one of the most popular subjects in the quantum information science. In this paper, we construct a \textbf{linear}-entropy-based non-Markovianity witness scheme. The positive definiteness of the Choi state will be broken in the non-Markovian evolution, which can be witnessed by its linear entropy. Thus, the \textbf{linear} entropy of the Choi state can be used to witness the non-Markovian dynamics. The effectiveness of the proposed method is verified by an example of the pure dephasing channel. Also, we show that this method can be extended to the one based on R\'{e}nyi entropy.
\end{abstract}
\shortabstract
\begin{document}
\maketitle

\section{\textbf{{Introduction}}}

As the most widely used physical platform in the quantum information science, open quantum system has been extensively investigated in recent years from both the fundamental and applicative perspective \cite{1,2,3}. The open systems, due to the inevitable interaction with the external environment, present some different dynamics phenomena, which are related to the strength of the system-environment coupling, structured reservoirs, low temperatures, and initial system-environment correlations \cite{4,5,6,7,8,9}. According to their dynamical features, the open quantum systems are divided into non-Markovian dynamics systems and Markovian dynamics systems. Non-Markovian dynamics are the ones with memory effects, which indicates a backflow of information from the environment to the system. In contrast, Markovian systems, the systems \textbf{without memory effects}, refer to the ones without the backflow of information from the environment to the system.

In general, the open system is firstly assumed to be Markovian in the theoretical investigations, because the Markovian dynamics has well defined mathematical structure \cite{9,10,11,12,12l}. However, Markovian assumption does not always apply to the physical systems of interest. For instance, Born-Markov approximation is violated when the system-environment coupling is not sufficiently weak, leading to the non-Markovian effects that cannot be neglected \cite{10,11,12}. Thus, the investigation of the non-Markovianity is necessary. Also, it has been found that non-Markovianity plays a leading role in the quantum information processing, such as quantum control \cite{13,14,15}, entangled states preparation \cite{16}, quantum metrology \cite{17} , quantum resource theory \cite{18}, and quantum evolution speedup \cite{19,20,21}. Therefore, non-Markovian evolutions of the open quantum systems \cite{22,23,24} \textbf{have drawn more and more interest} in both theory \cite{25,26,27} and experimentation \cite{28,29,30}.

Recently, various methods have been proposed from different points of view to detect non-Markovianity \cite{31,32,33,34,35,36,37,38,39}, for instance the one based on quantum uncertainty relations \cite{31,32}. The seminal and powerful work by Ref. \cite{32}, indicates that the uncertainty relation for the Choi state will be violated when the dynamics is non-Markovianity, thus the uncertainty relations can be used to detect non-Markovianity. In this paper, we construct a non-Markovianity witness scheme along the way of Ref. \cite{32}. We show that the linear entropy of the Choi state will be negative for the non-Markovian dynamics. Thus, the non-Markovianity can be witnessed by the negative linear entropy.

The outline of the paper is as follows. In Sec. 2, the concepts of Choi state and linear entropy are introduced. Sec. 3 is used to construct the linear-entropy-based non-Markovianity witness scheme. In Sec. 4, we illustrate the effectiveness of the scheme by the pure dephasing channel. Also, by making a comparison with the scheme based on the uncertainty relation, we show that our scheme is more effective in some cases. In Sec. 5, the scheme is extended to the one based on R\'{e}nyi entropy. The conclusion is presented in Sec. 6.

\section{\textbf{Preparations}}

This section is mainly used to introduce the concepts of Choi state and linear entropy, as well as their properties. Consider a quantum map in the $d$-dimensional Hilbert space $\mathcal{H}$:
\begin{align}\label{1}
\Lambda(t, t+\varepsilon) : \rho(t) \rightarrow \rho(t+\varepsilon) \quad \mathcal{H} \rightarrow \mathcal{H}\tag{1},
\end{align}
where $t$ represents time and $\varepsilon$ is a small-time interval. Assume that there exists another same Hilbert space $\mathcal{H}$, and $|\psi\rangle= 1 / \sqrt{d} \sum_{i=1}^{d}|i\rangle|i\rangle$ is the maximally entangled state of the Hilbert space $\mathcal{H} \otimes \mathcal{H}$, with $\{|i\rangle\}_{i=1}^{d}$ being an orthonormal basis of the Hilbert space $\mathcal{H}$. Then, the Choi state of the map $\Lambda$ is defined as \cite{40,41,42,43}:
\begin{align}\label{1}
\operatorname{Choi}(\Lambda)=[\mathbb{I} \otimes \Lambda(t, t+\varepsilon)](|\psi\rangle\langle\psi|)\tag{2},
\end{align}
where $\mathbb{I}$ is the identity map of the Hilbert space $\mathcal{H}$. Based on the definition of the Choi state, some properties of the Choi state can be obtained \cite{42,43}: (i) the Choi state $\operatorname{Choi}(\Lambda)$ is positive semi-definite when the map  $\Lambda$  is completely-positive (CP); and (ii) \textbf{the trace of $\operatorname{Choi}(\Lambda)$ state is 1} when the map $\Lambda$ is trace-preserving.

The linear entropy reads:
\begin{align}\label{1}
S_{l}(\rho)=\frac{d}{d-1}\{1-\operatorname{Tr}(\rho^{2})\}\tag{3},
\end{align}
where $\rho$ is density matrix of the system. We have $0 \leq S_{l}(\rho) \leq 1$, namely $S_{l}(\rho) \in[0,1]$, when $\operatorname{Tr}(\rho)=1$ and $\rho : \geq 0,$ with $\rho : \geq 0$ representing that $\rho$ is positive semi-definite.

\section{\textbf{Non-Markovianity Detection based on Linear Entropy}}

In this section, we construct a linear-entropy-based non-Markovianity witness scheme. Such a scheme provides a criterion for the non-Markovian dynamics, and also the effectiveness and the advantage of this criterion are demonstrated.

\textbf{Theorem:}  A quantum dynamics evolution $\Lambda$ is non-Markovian, when the linear entropy of the Choi state  $\operatorname{Choi}(\Lambda)$  is negative, namely $S_{l}(\operatorname{Choi}(\Lambda))<0$.

\noindent \emph{Proof: }Here we use the \emph{reductio and absurdum} to prove the theorem above. Assume that (a) the quantum dynamics evolution $\Lambda$ is Markovian, and (b) $S_{l}(\operatorname{Choi}(\Lambda))<0$.

Based on the definition of the Choi state, we have the $\operatorname{Choi}(\Lambda)$ is Hermitian \cite{32}. Then, the Choi state $\operatorname{Choi}(\Lambda)$ can be decomposed as:
\begin{align}\label{1}
\operatorname{Choi}(\Lambda)=\sum_{m} \lambda_{m} [\lambda_{m} ]\tag{4},
\end{align}
where $\lambda_{m}$ is the eigenvalue of the $\operatorname{Choi}(\Lambda)$, $|\lambda_{m}\rangle$ is the corresponding eigenstate, and the projection operator $[ \lambda_{m} ]=|\lambda_{m}\rangle\langle\lambda_{m}|$.

According to the assumption (a), one can obtain that the evolution $\Lambda$ is CP. Then, utilizing the prosperity (i) of the Choi state, one can obtain that the Choi state $\operatorname{Choi}(\Lambda)$ is positive semi-definite, namely:
\begin{align}\label{1}
\lambda_{m} \geq 0\tag{5}.
\end{align}
Also, based on the prosperity (ii) of the Choi state, we have:
\begin{align}\label{1}
\operatorname{Tr}[\operatorname{Choi}(\Lambda)]=\sum_{m} \lambda_{m}=1\tag{6}.
\end{align}
Combining (5) and (6), one can obtain:
\begin{align}\label{1}
0 \leq \lambda_{m} \leq 1\tag{7}.
\end{align}
Therefore, based on (7), we deduce:
\begin{align}\label{1}
S_{l}(\operatorname{Choi}(\Lambda))&=\frac{d}{d-1}(1-\operatorname{Tr}(\rho^{2}))\nonumber \\
&=\frac{d}{d-1}(1-\sum_{m} \lambda_{m}^{2}) \geq 0\tag{8}.
\end{align}
Obviously, the deduction of the assumption (a) is contradictory with the assumption (b). Then, we can obtain that the evolution $\Lambda$ is non-Markovian when $S_{l}(\operatorname{Choi}(\Lambda))<0$ has been detected. Thus, the negative linear entropy can be considered as a signature of the non-Markovianity.

In fact, the reason why the non-Markovianity is witnessed by the negative linear entropy can be explained as follows. The evolution $\Lambda$ is CP when $\Lambda$ is Markovian, and the CP definiteness will be broken, so do the positive definiteness of $\operatorname{Choi}(\Lambda)$, when the evolution $\Lambda$ is non-Markovian. Thus, the eigenvalue of Choi state can be negative for the non-Markovian evolution, which means Eq. (5) and Eq. (7) are violated. $S_{l}(\operatorname{Choi}(\Lambda))$ can be negative when Eq. (5) and Eq. (7) are violated, and thus, one can obtain that the evolution $\Lambda$ is non-Markovian when $S_{l}(\operatorname{Choi}(\Lambda))<0$ has been detected.

\section{\textbf{Dephasing channel as an Example}}

In this section, a pure dephasing channel will be used as an illustration to demonstrate the theoretical conclusions obtained above. Consider a qubit system coupled to a thermal reservoir, which is modeled by an infinite set of harmonic oscillators, and then the total Hamiltonian of the composite system can be written as \cite{33}:
\begin{align}\label{1}
H=\omega_{\sigma} \sigma_{z}+\sum_{k} \omega_{k} a_{k}^{\dagger} a_{k}+\sum_{k} g_{k} \sigma_{z}(e^{i \theta_{k}} a_{k}+e^{-i \theta_{k}} a_{k}^{\dagger})\tag{9},
\end{align}
where $\sigma_{z}$ is a Pauli operator of the qubit, $\omega_{\sigma}$ represents the energy gap of the qubit system, $\omega_{k}$ is the frequency of the k-th reservoir mode, $a_{k}(a_{k}^{\dagger})$  is the corresponding annihilation (creation) operators ,  \textbf{$g_{k}$ is the modulus of the coupling constant \cite{33}, and $\theta_{k}$ is the corresponding phase \cite{33}}. Assume the density operator of the qubit system is denoted by $\rho(t)$, and then the Lindblad master equation for $\rho(t)$ is given as \cite{33}:
\begin{align}\label{1}
\frac{d}{d t} \rho(t)=\gamma(t)(\sigma_{z} \rho(t) \sigma_{z}-\rho(t))\tag{10},
\end{align}
where $\gamma(t)$ reads:
\begin{align}\label{1}
\gamma(t)=\frac{2 \lambda \gamma_{0} \sinh (tg/2)}{g \cosh (tg/2)+\lambda \sinh (tg/2)}\tag{11},
\end{align}
with $g=\sqrt{\lambda^{2}-2 \gamma_{0} \lambda}$, $\lambda$ and $\gamma_{0}$ are two positive constants related to the thermal reservoir. $\gamma(t)$ is the time-dependent dephasing rate determined by the spectral density of the thermal reservoir \cite{32}, and the dynamics described by (10) is non-Markovian when $\gamma(t)$  is negative \cite{44}.

Taking advantage of the definition of the Choi state, one can obtain the Choi state for the master equation (10):
\begin{align}\label{1}
\mathcal{C}(\mathrm{t})=\left(\begin{array}{cccc}{\frac{1}{2}} & {0} & {0} & {\frac{1}{2} e^{\chi}} \\ {0} & {0} & {0} & {0} \\ {0} & {0} & {0} & {0} \\ {\frac{1}{2} e^{\chi}} & {0} & {0} & {\frac{1}{2}}\end{array}\right)\tag{12},
\end{align}
where $\chi=-4 \gamma_{0} \varepsilon \lambda /[\lambda+g \operatorname{coth}(t g / 2)]$. Then, the linear entropy of the Choi state $\mathcal{C}(\mathrm{t})$ can be obtained:
\begin{align}\label{1}
S_{l}(\mathcal{C}(\mathrm{t}))=\frac{1}{3}(2-2 e^{-8 \gamma_{0} \varepsilon \lambda /[\lambda+g \cosh (t g / 2)]})\tag{13}.
\end{align}
The time evolutions of the Lindblad coefficient $\gamma(t)$ and the linear entropy $S_{l}(\mathcal{C}(\mathrm{t}))$ are demonstrated in Fig. 1. From Fig. 1, we can see that the linear entropy $S_{l}(\mathcal{C}(\mathrm{t}))$ is negative only when the Lindblad coefficient $\gamma(t)$ is negative. As mentioned above, the dynamics is non-Markovian when $\gamma(t)$ is negative \cite{44}. Thus, the non-Markovian dynamics can be effectively witnessed by the negative linear entropy.

\begin{figure}
\centering 
\includegraphics[height=4.1cm]{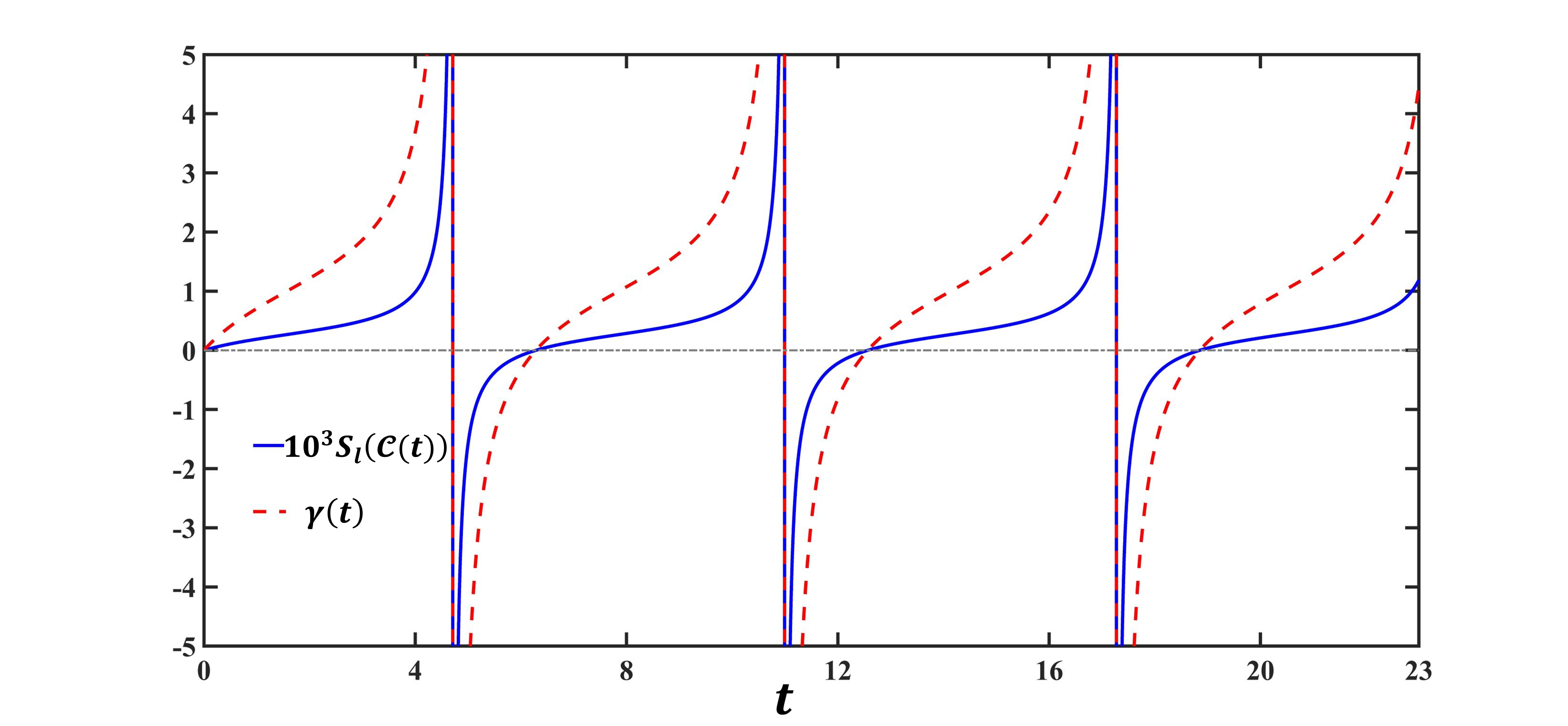}
\caption{Evolution of the linear entropy $S_{l}(\mathcal{C}(\mathrm{t})))$ and $\gamma(t)$ with respect to time $t$. The related parameters are taken as
$\gamma_{0}=1$, $\lambda=1$ and $\varepsilon=1 \times 10^{-4}$. The dynamics exhibits non-Markovianity, when $\gamma(t)$ is negative. We can see that non-Markovian dynamics can be effectively witnessed by the negative linear entropy $S_{l}(\mathcal{C}(\mathrm{t}))$. To make a clearer comparison between  $S_{l}(\mathcal{C}(\mathrm{t}))$) and $\gamma(t)$ , we here plot $10^3*S_{l}(\mathcal{C}(\mathrm{t}))$.
} 
\label{s1}
\end{figure}

The seminal and powerful work Ref. \cite{32} firstly deduced that the dynamics is non-Markovian when the uncertainty relation for the Choi state is violated, and thus the uncertainty relations can be used to detect non-Markovianity. Here, taking the uncertainty-relation-based non-Markovianity witness scheme as a comparison, we show that our scheme is more effective in some cases.

The most famous uncertainty relation is the Schr\"{o}dinger-Robertson uncertainty relation (SUR) \cite{45}, which reads:
\begin{align}\label{9}
{\Delta A}^{2}{\Delta B}^{2} \geq \frac{1}{4}|\langle[A, B]\rangle|^{2}+\frac{1}{4}|\langle\{\check{A}, \check{B}\}\rangle|^{2}\tag{14},
\end{align}
where $A$ and $B$ stand for two arbitrary incompatible observables. $\Delta O^{2}$ represents the variance of the observable $O$, $\langle O\rangle$ is the expectation of $O$, and $\check{O}=O-\langle O\rangle$. $[A, B]=A B-B A$ and $\{\check{A}, \check{B}\}=\check{A} \check{B}+\check{B} \check{A}$ are the commutator and anti-commutator, respectively. Also, SUR can be reformed as the sum form:
\begin{align}\label{9}
{\Delta A}^{2}+{\Delta B}^{2} \geq  \sqrt{|\langle[A, B]\rangle|^{2}+|\langle\{\check{A}, \breve{B}\}\rangle|^{2}}\tag{15},
\end{align}
where the inequality $\Delta A^{2}+\Delta B^{2} \geq 2 \Delta A \Delta B$ is used. Ref. \cite{32} deduced that the sum form uncertainty relation is more effective to detect the non-Markovianity than the product form uncertainty relations in some cases. Thus, in the following, we mainly focus on the scheme based on the sum form uncertainty relation (15). In order to witness the non-Markovianity by the uncertainty relation (15), we define the following quantity:
\begin{align}\label{9}
Q(\mathrm{A}, \mathrm{B}, \rho)={\Delta A}^{2}+{\Delta B}^{2} - \sqrt{|\langle[A, B]\rangle|^{2}+|\langle\{\check{A}, \check{B}\}\rangle|^{2}}\tag{16},
\end{align}
where the variance and the expectation are calculated over the state $\rho$. The uncertainty relation (15) is violated on the state $\rho$ when $Q(A, B, \rho)$ is less than zero, and thus the negative $Q(A, B, \rho)$ can be considered as a signature of non-Markovianity. For the Choi state of evolution (10), one can obtain \cite{46}:
\begin{align}\label{9}
Q(A, B, \mathcal{C}(t))=\frac{5}{4}-e^{-8 \gamma_{0} \varepsilon \lambda /[\lambda+g \cosh (tg/2)]}\tag{17}.
\end{align}
The time evolutions of $Q(A, B, \mathcal{C}(t))$, $S_{l}(\mathcal{C}(\mathrm{t}))$ and $\gamma(t)$ are presented in Fig. 2.

\begin{figure}
\centering 
\includegraphics[height=4.2cm]{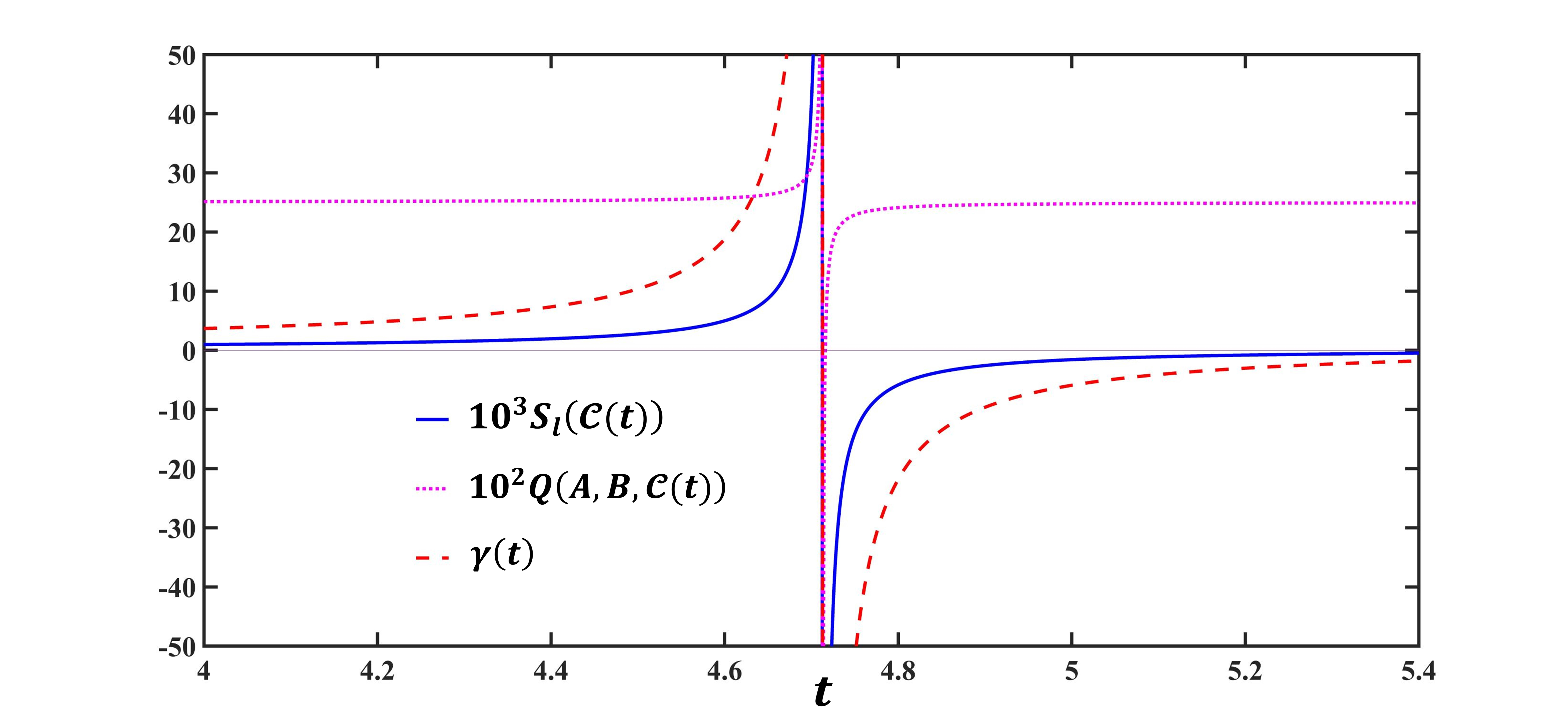}
\caption{Evolutions of $Q(A, B, \mathcal{C}(t))$, $S_{l}(\mathcal{C}(\mathrm{t}))$ and $\gamma(t)$ with respect to time are presented. Here, the related parameters are taken as $\gamma_{0}=1$, $\lambda=1$ and $\varepsilon=1 \times 10^{-4}$. To make a clearer comparison between  $Q(A, B, \mathcal{C}(t))$, $S_{l}(\mathcal{C}(\mathrm{t}))$ and $\gamma(t)$, we here plot $10^3*S_{l}(\mathcal{C}(\mathrm{t}))$ and $10^2*Q(A, B, \mathcal{C}(t))$.
} 
\label{s2}
\end{figure}

As we can see from Fig.2, the linear-entropy-based scheme is more effective to detect non-Markovianity than uncertainty-relation-based scheme. For instance, the dynamics of the evolution is non-Markovian when $t=4.8$, since $\gamma(t)<0$ for $t=4.8$. From Fig.2, one can see that $S_{l}(\mathcal{C}(\mathrm{t}))<0$ and $Q(A, B, \mathcal{C}(t))>0$ around the moment $t=4.8$.  Thus, the non-Markovianity at this moment can only be witnessed by the linear-entropy-based scheme.

\textbf{ In fact, the information of non-Markovianity is mainly reflected in all the eigenvalues of the Choi state, and, according to the definition of the linear entropy, we can see that the linear  entropy of Choi state contains the information of all eigenvalues of the Choi state. Thus, the linear entropy can be well used to capture the information of the non-Markovianity. To demonstrate this point, we make a comparison between our method and a well-defined quantum non-Markovianity measure \cite{10}, which is used to qualify the extent of the non-Markovianity. As shown in Fig.3, we can see our witness scheme has a very similar evolution to the standard quantum non-Markovianity measure, which means the linear entropy is well used to capture the information of the non-Markovianity.}

\begin{figure}
\centering 
\includegraphics[height=4.2cm]{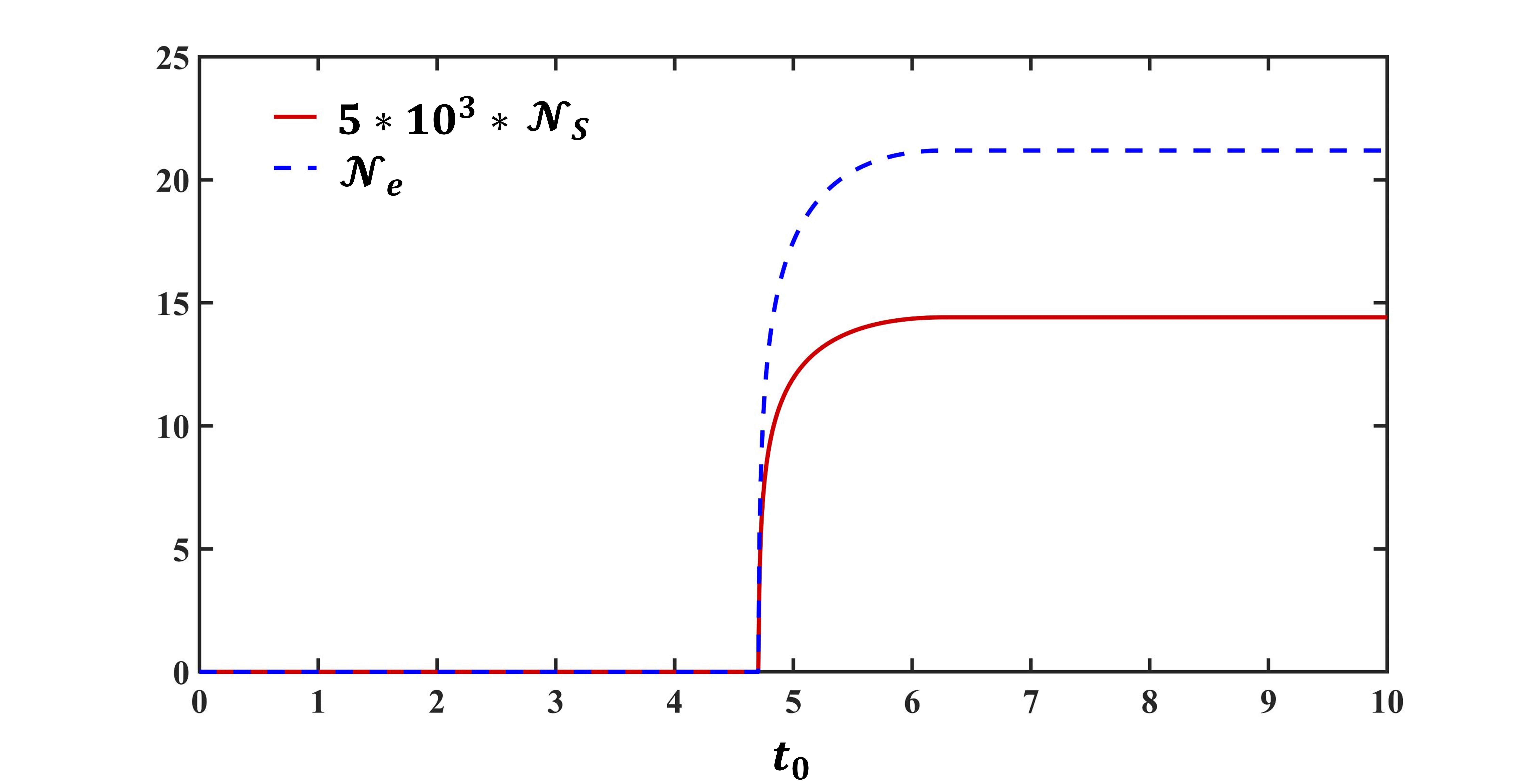}
\caption{\textbf{Comparison between our method and the standard non-Markovianity measure $\mathcal{N}_e$  provided in Ref.\cite{10}. The $\mathcal{N}_e$ is used to quantify the extent of the non-Markovianity, thus it can capture the full knowledge of the non-Markovianity. As mentioned in the main text, the dynamics is non-Markovian when $S_{l}(Choi(\Lambda))$ is less than zero. Thus, we can use $\mathcal{N}_S=-\int^{t_0}_{0,S_{l}(Choi(\Lambda))\leq0}S_{l}(Choi(\Lambda))dt$ to qualify the information of the non-Markovianity containing in the $S_{l}(Choi(\Lambda))$ during the time period $[0,t_0]$. The figure illustrates the comparison between $\mathcal{N}_S$ and $\mathcal{N}_e$ in the dephasing channel, where $\mathcal{N}_{e}=-2 \int_{0 \gamma(\mathrm{t}) \leq 0}^{t_{0}} \gamma(\mathrm{t}) d t$. We can see that they have the very similar evolution.}
} 
\end{figure}

\section{\textbf{Extention to R\'{e}nyi Entropy}}

The R\'{e}nyi entropy has been widely used in quantum information processing, and it is an important tool to solve various problems in the quantum information theory, such as entanglement characterization \cite{47}. The $\alpha$-order quantum R\'{e}nyi entropy reads \cite{48}:
\begin{align}\label{9}
S_{\alpha}(\rho)=\frac{1}{1-\alpha} \log _{2} \operatorname{Tr}(\rho^{\alpha})\tag{18},
\end{align}
where $\alpha \in(0,1) \cup(1, \infty)$. R\'{e}nyi entropy for different order will recover some important quantum measures. For instance, the R\'{e}nyi entropy reduces to von Neumann entropy in the limit $\alpha \rightarrow 1$, minimum entropy in the limit $\alpha \rightarrow \infty$, and skew information for $\alpha=1 / 2$ . In particular, R\'{e}nyi entropy is an alternative form of linear entropy when $\alpha=2$.

Then, we show that the non-Markovianity can also be detected by the R\'{e}nyi entropy. Assume that (1) the dynamics of the evolution $\Lambda$ is Markovian and (2) the R\'{e}nyi entropy of the Choi state $\operatorname{Choi}(\Lambda)$ is negative. Based on the assumption (1), one can obtain $0 \leq \lambda_{\mathrm{m}} \leq 1$, and thus $\log _{2} \operatorname{Tr}(\rho^{\alpha}) /(1-\alpha)=\log _{2}(\sum_{m} \lambda_{m}^{\alpha}) /(1-\alpha) \geq 0$ for $(0,1) \cup(1, \infty)$. Obviously, the deduction based on the assumption (1) is contradictory with the assumption (2). Thus, one can obtain that the dynamics is non-Markovian when the R\'{e}nyi entropy of the corresponding Choi state is negative. That is to say, similar to the negative linear entropy, the negative R\'{e}nyi entropy can also be considered as a signature of the non-Markovianity, as shown in Fig.4.
\begin{figure}
\centering 
\includegraphics[height=4.2cm]{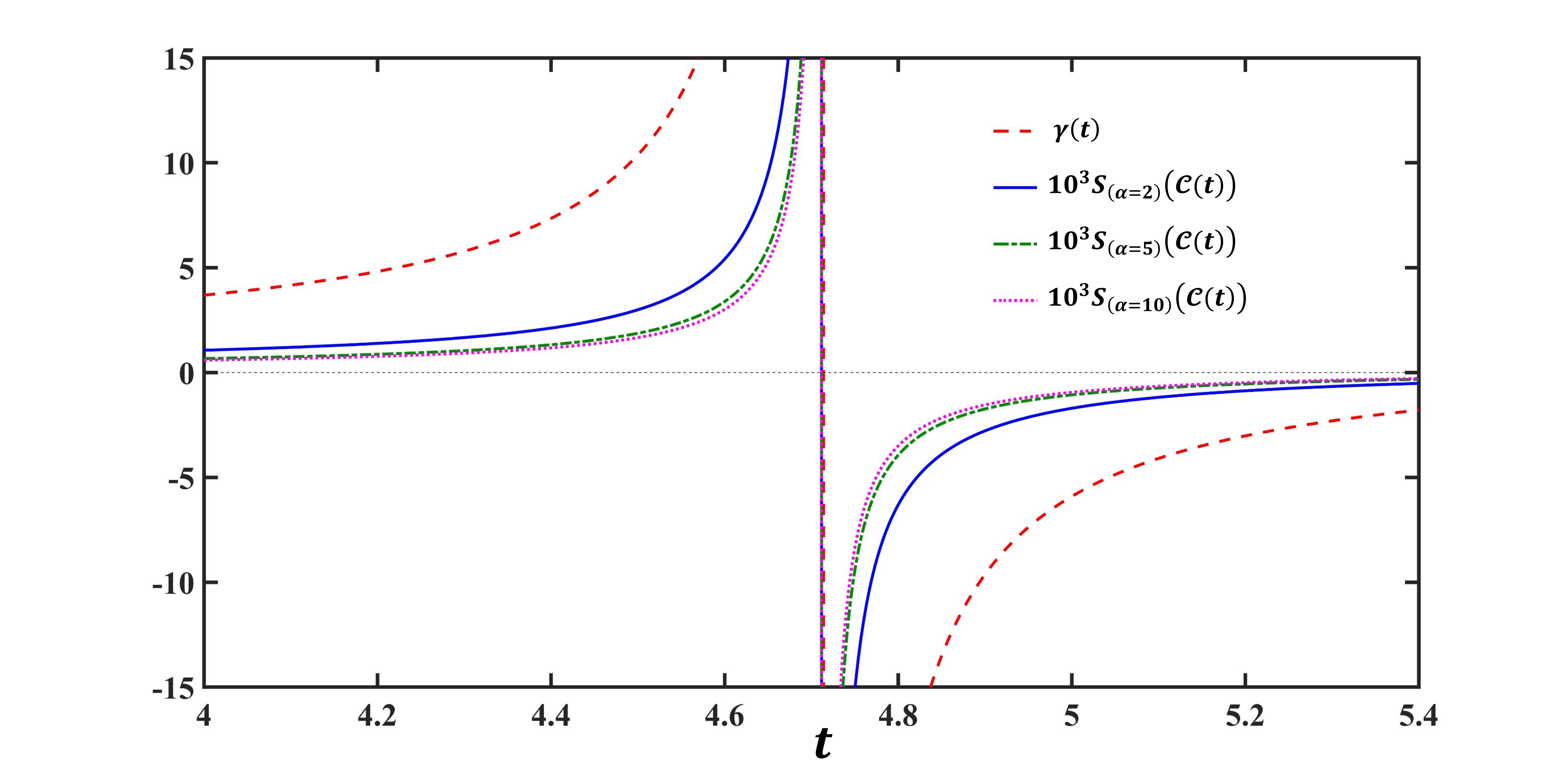}
\caption{Evolutions of $\gamma(t)$, $S_{(\alpha=2)}(\mathcal{C}(t))$, $S_{(\alpha=5)}(\mathcal{C}(t))$, and $S_{(\alpha=10)}(\mathcal{C}(t))$ with respect to time are presented. Here, the related parameters are taken as $\gamma_{0}=1$, $\lambda=1$ and $\varepsilon=1 \times 10^{-4}$. To make a clearer comparison between  $\gamma(t)$, $S_{(\alpha=2)}(\mathcal{C}(t))$, $S_{(\alpha=5)}(\mathcal{C}(t))$, and $S_{(\alpha=10)}(\mathcal{C}(t))$, we here plot $10^3*S_{(\alpha=2)}(\mathcal{C}(t))$, $10^3*S_{(\alpha=5)}(\mathcal{C}(t))$, and $10^3*S_{(\alpha=10)}(\mathcal{C}(t))$.
} 
\label{s2}
\end{figure}

\section{\textbf{{Conclusion}}}
In conclusion, we propose a new method to detect non-Markovianity by the \textbf{linear} entropy of the Choi state. The positivity of the Choi state will be broken for the non-Markovian dynamics, and the negative Choi state can be witnessed by the \textbf{linear} entropy. Thus, the \textbf{linear} entropy can be used to detect the non-Markovianity. The effectiveness of this scheme is illustrated by the pure dissipative channel. Also, we show that the scheme we constructed is more effective than the uncertainty-relation-based non-Markovianity witness scheme in some cases. Finally, the \textbf{linear}-entropy-based non-Markovianity witness scheme is extended to the one based on the R\'{e}nyi entropy.

\section*{Acknowledgements}
This work was supported by NSFC under grants No. 11574022.

\section*{Conflict of Interest}
The authors declare no conflict of interest.

\end{document}